\documentclass[twocolumn,preprintnumbers,prl,letterpaper,superscriptaddress,showpacs]{revtex4}
\usepackage{amsmath}
\usepackage{graphicx}
\usepackage{dcolumn}
\usepackage{bm}
\usepackage{bbold}
\usepackage{subfigure}

\begin{document}

\title{Dynamic Stimulation of Quantum Coherence in Systems of Lattice Bosons}
\author{Andrew Robertson}
\affiliation{Joint Quantum Institute and Condensed Matter Theory Center, Department of
Physics, University of Maryland, College Park, Maryland 20742-4111, USA}
\author{Victor M. Galitski}
\affiliation{Joint Quantum Institute and Condensed Matter Theory Center, Department of
Physics, University of Maryland, College Park, Maryland 20742-4111, USA}
\author{Gil Refael}
\affiliation{Department of Physics, California Institute of Technology, 1200 E. California Boulevard, Pasadena, California 91125, USA}

\begin{abstract}
Thermal fluctuations tend to destroy long-range phase correlations. Consequently, bosons in a lattice will undergo a transition from a phase-coherent superfluid as the temperature rises. Contrary to common intuition, however, we show that non-equilibrium driving can be used to reverse this thermal decoherence. This is possible because the energy distribution at equilibrium is rarely optimal for the manifestation of a given quantum property. We demonstrate this in the Bose-Hubbard model by calculating the non-equilibrium spatial correlation function with periodic driving. We show that the non-equilibrium phase boundary between coherent and incoherent states at finite bath temperatures can be made qualitatively identical to the familiar zero-temperature phase diagram, and we discuss the experimental manifestation of this phenomenon in cold atoms.
\end{abstract}

\pacs{64.60.Ht, 03.75.Lm, 03.75.Kk}
\date{\today }
\maketitle

A system of bosons confined to a lattice has long represented an alluring opportunity to study the interplay between two phenomena at the heart of the academic and industrial interest in many-body quantum mechanics: particle tunneling and phase coherence. The Bose-Hubbard model (BHM) describing such systems in the tight-binding approximation is much richer than its simple mathematical form betrays. It admits such novelties as dynamic localization \cite{Zenesini10,Creffield06}, photon-assisted tunneling \cite{Weiss09,Eckardt05}, as well as an archetypal example of a quantum phase transition between superfluid and insulator-like states \cite{Freericks94,Fisher89}. Thermal fluctuations destroy these quantum effects, but it is comparatively unknown that deliberately driving the system out of equilibrium can moderate or reverse entirely the destructive effect of raising the temperature.

Despite its obscurity, it has been known since the 1960's that pushing a system out of equilibrium can enhance its quantum properties. In 1966, Wyatt et. al. \cite{Wyatt66} showed that illuminating a microbridge could stimulate its superconductivity. Eliashberg explained this result in 1970 by calculating the nonequilibrium quasiparticle distribution induced by the radiation \cite{Eliashberg70}. Blamire et. al. later demonstrated that superconducting transition temperatures could be enhanced in this manner by several times their equilibrium values \cite{Blamire91}. More recently, the idea of non-equilibrium phase transitions has garnered interest in studies of optically trapped atoms \cite{Robertson09} and induced topological structure \cite{Lindner11}.

The reason for enhancement goes as follows. The quantum properties of a system depend on the energy distribution of excitations. It is easily verified that the equilibrium distribution is rarely optimal for the enhancement of a chosen property. A brief survey of the model at equilibrium reveals this to be the case for the BHM. Indeed, an analogue of photon-assisted tunneling (PAT) \cite{Creffield06,Eckardt05} has already been theorized for the BHM at $T=0$. However, a fully non-equilibrium treatment including the effects of temperature (and how they may be mitigated) is not known to us.

In this Letter, we shall show that harmonically driving a system of lattice bosons connected to a thermal reservoir can increase the region in parameter space where the quantum coherent phase exists. Even for finite temperatures of the bath, the phase diagram of the BHM can be made qualitatively identical to the $T=0$ diagram. We shall demonstrate this by defining non-equilibrium correlation functions $\langle a^{\dagger}_{i}(t) a_{j}(t^{\prime}) \rangle$ within the Keldysh and Floquet formalisms \cite{Kamenev09,Keldysh64,Mahan81,Brandes97,Tsuji08}. Divergence of the real part of this quantity corresponds to infinite long-range correlations. This will define the phase boundary between superfluid and incoherent states for our non-equilibrium system. We shall find these functions perturbatively \cite{Metzner91,Ohliger08} in the small quantity $J/U$ and arrive at a Dyson equation that has both ordinary and entry-wise (Hadamard) matrix products. This novel structure will then be solved by column vectorization for the stationary non-equilibrium correlation function. 

To see how superfluidity may be enhanced by means of a non-equilibrium pulse, let us briefly review the BHM at equilibrium. The hamiltonian of the BHM is
\begin{equation}
H_{0}+H_{J}=\frac{1}{2} U \sum_{i}{a^{\dagger}_{i} a_{i} \left(a^{\dagger}_{i} a_{i}-1\right)}-\mu \sum_{i}{a^{\dagger}_{i} a_{i}}-\sum_{ij}{J_{ij} a^{\dagger}_{i} a_{j}}
, \label{HBHM}
\end{equation}
where nearest neighbor tunneling of strength $J_{ij}=J$ is assumed on a 2D square lattice with $T \ll U$. This bath temperature models the coupling to an environment \cite{Dalidovich09} that dissipates energy. Let us first consider the $T=0$ case where $\mu / U$ is close to some integer $M$ so that $\mu / U=M \pm \delta$ for some $0 < \delta < 1/2$. Letting the tunneling $J$ be infinitesimally small, the ground state is a Mott insulator. The energy gap for adding a particle or hole is proportional to $\delta$ and ($1- \delta$) respectively. As $\delta$ is tuned to zero (unity), the state with an extra particle (hole) becomes degenerate with the Mott insulator. Thus, excess particles(holes) will be free to hop among sites with no energy barrier. At low temperatures, they will condense producing superfluidity \cite{Fisher89}. As $J$ is increased, the low lying excitations become long-range collective particle (hole) tunneling events between the system and reservoir. These events promote particle fluctuations, but they tend to stabilize the phase across sites through the number-phase conjugate relationship. The manifold where the energies needed for these long-range excitations vanishes defines the phase boundary \cite{Freericks94} at $T=0$. When $T$ is finite, sites will have a thermal probability $p_{n} = e^{- \beta \varepsilon_{n}}$ of being occupied by $n$ bosons. Sites will have different energies, and their phases will rotate at different rates. This aggravates the phase fluctuations that destroy superfluidity. The region in $J$ vs. $\mu$ space that permits superfluidity is thus reduced (especially near integer values of $\mu / U$) as the temperature is increased. We will see that the long-range phase-coherence demonstrated by our non-equilibrium correlation functions can be interpreted physically as the artificial closing of the Mott gap thereby allowing for the excitation of long-range tunneling modes.



To effect an enhancement of superfluidity, we shall need a perturbation that pushes the system out of equilibrium. A thermal bath is also necessary both to counterbalance the heating induced by the perturbation as well as to ensure that the concept of temperature remains well-defined. We shall add three terms to the Hamiltonian so that $H \left( t \right)=H_{0}+H_{J}+H_{\text{bath}}+H_{\text{coup}}+H_{V} \left(t \right)$. The bath and coupling Hamiltonians $H_{\text{bath}}=\sum_{i \alpha}{\varepsilon_{\alpha} b^{\dagger}_{i \alpha} b_{i \alpha}}$ and $H_{\text{coup}}=\sum_{i \alpha}{g_{\alpha} \left(b^{\dagger}_{i \alpha} + b_{i \alpha} \right) a^{\dagger}_{i} a_{i} }$ model the coupling of each site to an infinite bath of oscillator degrees of freedom such as the collective modes of a larger condensate in the which the lattice is immersed. Since the only crucial function of the bath is to balance the energy input from the driving, many types of dissipation are possible and the results of this paper are likely to extend to these alternate models. While future work is needed to verify this assertion, low frequency irradiation of a Josephson array should easily corroborate or deny this intuition experimentally. We shall model the strength of our bath coupling \cite{Dalidovich09} by a purely local and Ohmic parameter $g_{\alpha}^{2}=\eta \varepsilon_{\alpha} \exp \left\{-\varepsilon_{\alpha}/\Lambda \right\}$. The driving term that will force a departure from equilibrium is given by
\begin{equation}
H_{V} \left(t \right)=V \sum_{i}{a^{\dagger}_{i} a_{i} \cos \left(\textbf{k} \cdot \textbf{x}_{i} - \Omega t \right)}
, \label{NEQH}
\end{equation}
In practice, $H_{V} \left( t \right)$ describes what is called a Bragg pulse \cite{Robertson09,Blakie02,Rey05}. In the limit where energy differences between internal atomic levels are much larger than $U$, $J$, and $T$, the potential in Eq.\ (\ref{NEQH}) is created by the superposition of two lasers offset to each other in both frequency and wave vector. The spatial dependence of our perturbation is necessary for nontrivial results because a constant perturbation simply multiplies the correlation function by a spatially constant phase factor \cite{Brandes97}. This factor is irrelevant to the tunneling of particles, and it can be gauged away by a time-dependent transformation of our operators. The precise form of the spatial dependence in our system, however, seems not to be very important and other schemes are certainly possible \cite{Creffield06}. We have chosen this one because of its experimental simplicity.

It is difficult to say much about the nature of the incoherent and coherent nonequilibrium phases or how they relate to the Kosterlitz-Thouless transition. Our focus will be to determine the boundary between the two non-equilibrium phases. To do this we shall approximate the correlation function $\langle a^{\dagger}_{i}(t) a_{j}(t^{\prime}) \rangle$ and look for divergences of this quantity over long distances. We shall do this perturbatively in the small quantity $J/U$ following the general method in \cite{Metzner91,Ohliger08}. Anticipating a nonequilibrium formalism and considering only the first order in the self-energy, the correlation function can be written as an infinite sum of simple chain diagrams defined on the foward-backward Keldysh contour, $C$. The evolution on $C$ is important in nonequilibrium problems because it dispenses with the need to know the state of the system at $t= \infty$ for the calculation of expectation values. The contour-time-ordering is accounted for by allowing green's functions to have a matrix structure \cite{Kamenev09,Mahan81}. That is, if we define $\hat{G} =
\begin{pmatrix}
G^{A} & 0 \\
G^{K} & G^{R}
\end{pmatrix} $ where $A,K,R$ refer to advanced, Keldysh, and retarded Green's functions, then the non-equilibrium Dyson series can be written as a sum of matrix products of correlation functions.
\begin{equation}
\hat{G}_{ij} \left(t,t^{\prime} \right) = \hat{g}_{ij} \left(t,t^{\prime} \right) - \sum_{i_{1} i_{1}^{\prime}}{J_{i_{1} i_{1}^{\prime}} \int_{- \infty}^{\infty} dt_{1} \hspace{5pt} \hat{g}_{i i_{1}^{\prime}} \left(t,t_{1} \right) \hat{G}_{i_{1} j} \left(t_{1},t^{\prime} \right) } 
, \label{KDys}
\end{equation}
where $\hat{g}_{ij}$ refers to the correlation functions with respect to the  hamiltonians $H_{0}+H_{\text{bath}}+H_{\text{coup}}+H_{V} \left(t \right)$ at bath temperature $T$. Because these hamiltonians are just sums of single-site terms proportional to products of density operators $n_{i}=a^{\dagger}_{i} a_{i}$, they are easy to diagonalize in the occupation basis. Additionally, the bath can be decoupled from the system \cite{Dalidovich09,Mahan81} via a canonical transformation  $a_{i} \rightarrow e^{S} a_{i} e^{-S}= a_{i} e^{ \sum_{\alpha}{\frac{g_{\alpha}}{\varepsilon_{\alpha}} \left( b^{\dagger}_{i \alpha} - b_{i \alpha} \right)}}= a_{i} X_{i}$ that uses the time-derivatives of the transformed fields to cancel the coupling to the bath. Averages of the form $\langle T_{c} a^{\dagger}_{i} \left(t \right) a_{j} \left(t^{\prime} \right) \rangle$ simply transform to $\langle T_{c} a^{\dagger}_{i} \left(t \right) a_{j} \left(t^{\prime} \right) \rangle \langle T_{c} X^{\dagger}_{i} \left(t \right) X_{j} \left(t^{\prime} \right) \rangle = \hat{g}_{ij} \left(t-t^{\prime} \right) \circ \hat{f}_{ij} \left(t-t^{\prime} \right)$  where $\circ$ denotes a Hadamard or entrywise product given by $\left( \hat{A} \circ \hat{B} \right)_{\alpha \beta} = A_{\alpha \beta} B_{\alpha \beta}$ where $\alpha$,$\beta$ designate components in the $2 \times 2$ Keldysh space. The inclusion of the driving field $H_{V} \left(t \right)$ produces a simple phase factor \cite{Brandes97} equal to $e^{i \frac{V}{\Omega} \left[ \sin \left(\textbf{k} \cdot \textbf{x}_{i} - \Omega t \right) - \sin \left(\textbf{k} \cdot \textbf{x}_{i} - \Omega t^{\prime} \right) \right] }$. We conclude that the non-equilibrium function $\hat{g}_{ij}$ is a product of the time-dependent factors produced by $H_{\text{bath}}+H_{\text{coup}}+H_{V} \left(t \right)$ and the bare function with respect to $H_{0}$. Expanding the non-equilibrium prefactor in terms of Bessel functions, the exact expression for $\hat{g}_{ij}$ is
\begin{eqnarray}
\hat{g}_{ij} \left(t,t^{\prime} \right)&=& \hat{g}^{\text{bare}}_{ij} \left( 
t - t^{\prime} \right) \circ \hat{f} \left(t-t^{\prime} \right) \nonumber \\
&\times& \hspace{-10pt} \sum_{m n = -\infty}^{\infty} \left(-1 \right)^{m+n} \mathcal{J}_{n} \left(\frac{V}{\Omega} \right) \nonumber \\
&\times& \mathcal{J}_{m} \left(- \frac{V}{\Omega} \right) e^{i \left[n \Omega t + m \Omega t^{\prime} - \left( n + m \right) \textbf{k} \cdot \textbf{x}_{i} \right]}
. \label{Hsub}
\end{eqnarray}
The functions $\hat{f}$ and $\hat{g}^{\text{bare}}_{ij}$ are $2 \times 2$ matrices of the equilibrium correlation functions for the bath and the system given by $H_{0}$ respectively. As they are equilibrium functions, they depend only on the difference of their time-arguments. All of the non-equilibrium information is stored in the expansion of the phase prefactor which depends on $t$ and $t^{\prime}$ separately. This is the source of the $\mathcal{J}_{0} \left( \frac{V}{\Omega} \right)$ dependence of the tunneling renormalization familiar from studies of dynamic localization \cite{Creffield06}.

The non-equilibrium phase factor makes $\hat{G}_{ij}$ a function of $\tau = \left( t+t^{\prime} \right) / 2$ rather than simply of $\Delta=t-t^{\prime}$. However, due to the time-periodicity of Eq.\ (\ref{NEQH}), $\hat{G}_{ij} \left(\tau,\Delta \right)$ is a function of $\tau$ only up to period $2 \pi / \Omega$. This discrete time symmetry of the Hamiltonian allows us to decompose every matrix function in Eq.\ (\ref{KDys}) as $\hat{G}_{ij} \left( t,t^{\prime} \right) = \frac{1}{2 \pi} \sum_{N}{e^{i N \Omega \tau} \int_{-\infty}^{\infty} e^{-i \omega \Delta} \hat{G}_{ij} \left( \omega,N \right)}$. Following the technique illustrated in \cite{Brandes97}, equation (\ref{KDys}) can now be written in terms of the functions $\hat{g}_{ij} \left( \omega,N \right)$, but it will be burdensome to work with it because the equation for $\hat{G}_{ij} \left( \omega, N \right)$ will include contributions from $\hat{G}_{ij} \left( \omega, N^{\prime} \right)$ for all $N^{\prime}$. Fortunately, we can mathematically represent this coupling as simple matrix multiplication if we transform to the so-called Floquet representation \cite{Tsuji08,Brandes97} defined as $\hat{G}_{ij} \left( \omega \right)_{mn} = \hat{G}_{ij} \left( \omega + \frac{m+n}{2} \Omega,m-n \right)$. We may now think of $\hat{G}_{ij} \left( \omega \right)_{mn}$ as an infinite square matrix in the two-dimensional space of Floquet indices $m$ and $n$. Each element of this matrix is itself a $2 \times 2$ Keldysh matrix. We will suppress the site indices and make use of the discrete translational symmetry of the problem by transforming to lattice-momentum space $\hat{G}_{i} \left( \omega, \textbf{q} \right)_{mn} = \sum_{j}{e^{i \textbf{q}_{i} \cdot \left( \textbf{x}_{j}-\textbf{x}_{i} \right)} \hat{G}_{ij} \left( \omega \right)_{mn}}$. Finally, we explicitly account for the plane-wave contribution to the full Green's function by defining $e^{i \left(n-m \right) \textbf{k}_{i} \cdot \textbf{x}_{i}} \hat{\mathcal{G}} \left( \omega, \textbf{q} \right)_{mn} = \hat{G}_{i} \left( \omega, \textbf{q} \right)_{mn}$. Having made these transformations, we arrive at an extremely simple form for the nonequilibrium Dyson equation.
\begin{equation}
\hat{\mathcal{G}} \left( \omega, \textbf{q} \right) = \hat{g} \left(\omega \right) \left[1-J \left( \textbf{q}, \textbf{k} \right) \circ \hat{\mathcal{G}} \left( \omega, \textbf{q} \right) \right]
, \label{NQDys}
\end{equation}
where for given matrices $A$ and $B$ in the Floquet space (indexed by $m,n$), $AB$ indicates an ordinary matrix product while $A \circ B$ denotes a Hadamard in the Floquet space rather than Keldysh space. The matrix $J \left( \textbf{q}, \textbf{k} \right)$ is a generalized lattice dispersion in Floquet space given by $J_{mn} \left( \textbf{q} \right) = J \sum_{\nu}{\cos \left[q_{\nu}+\left(m-n \right) k_{\nu} \right]}$ where $\nu=1,2$ denotes principle directions in our square lattice.

The existence of a Hadamard product in Eq.\ (\ref{NQDys}) complicates matters. We cannot merely multiply by inverses to solve for $\hat{\mathcal{G}}$ because there are now two types of inverses corresponding to the two types of products. This double-product structure in a Dyson equation seems so far unknown in any other context, but the situation can be salvaged by column vectorization (CV): the mapping of matrix $A$ to a vector $\vec{A}$ consisting of the first column of $A$ stacked on the next column and so on. We can then make use of the convenient identities relating Hadamard and ordinary matrix products through CV to solve Eq.\ (\ref{NQDys}) and rewrite it as
\begin{equation}
\vec{\mathcal{G}} \left( \omega, \textbf{q} \right) = \left( \mathbf{1}^{n_{p}^{2} \times n_{p}^{2}} - \left[ \mathbf{1}^{n_{p} \times n_{p}} \otimes  \hat{g} \left( \omega \right) \right] D \left\{ \vec{J} \left( \textbf{q} \right) \right\} \right)^{-1} \vec{g} \left( \omega \right)
, \label{Dvec}
\end{equation}
where $\otimes$ indicates a Kronecker product, and $D \left\{ \vec{A} \right\}$ denotes the diagonal matrix with entries given by those of $\vec{A}$. The identity matrix of size $k \times k$ is given by $\mathbf{1}^{k \times k}$, while $n_{p}$ is the size of the matrix $\hat{g}$ in the Floqet ($m,n$) space. It signifies how many higher harmonics we wish to include, or equivalently, the maximum time-resolution of our treatment. If we needed infinite time-resolution, we would of course let $n_{p}\rightarrow\infty$. However, each off-diagonal element $(m,n)$ will be weighted by $(J/U)^{m-n}$ while $\hat{g}_{mn} \rightarrow 0$ with increasing $m+n$, so we expect that $n_{p}$ need not be large to capture the relevant stationary behavior.

\begin{figure}[t]
\hspace{-0.25in}
\subfigure[]{\includegraphics[width=1.1in]{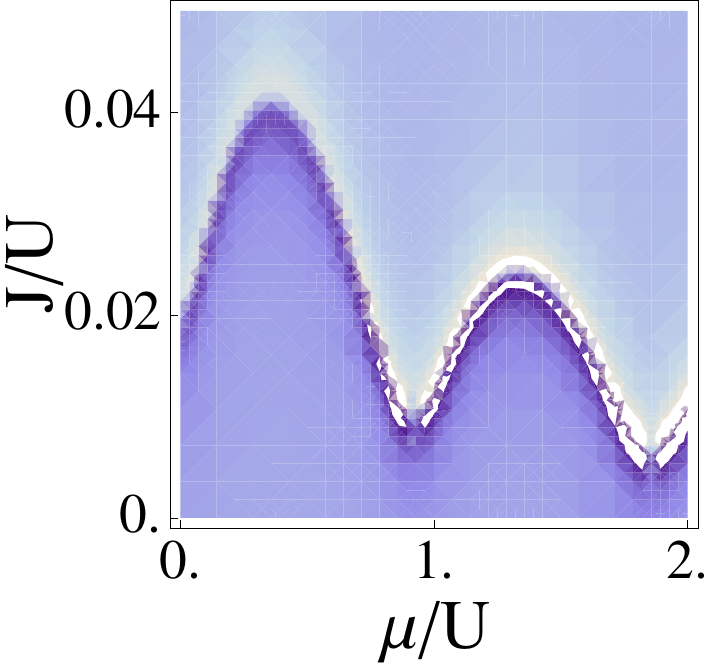}\label{fig:PDEQ}}
\hspace{0.01in}
\subfigure[]{\includegraphics[width=1.1in]{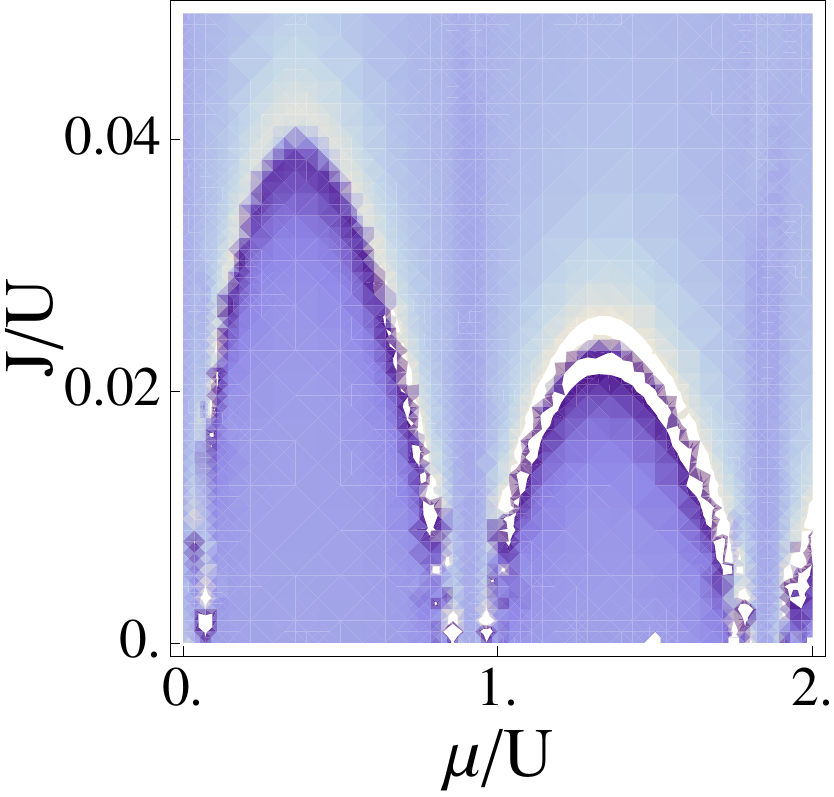}\label{fig:NQ05}}
\hspace{0.01in}
\subfigure[]{\includegraphics[width=1.1in]{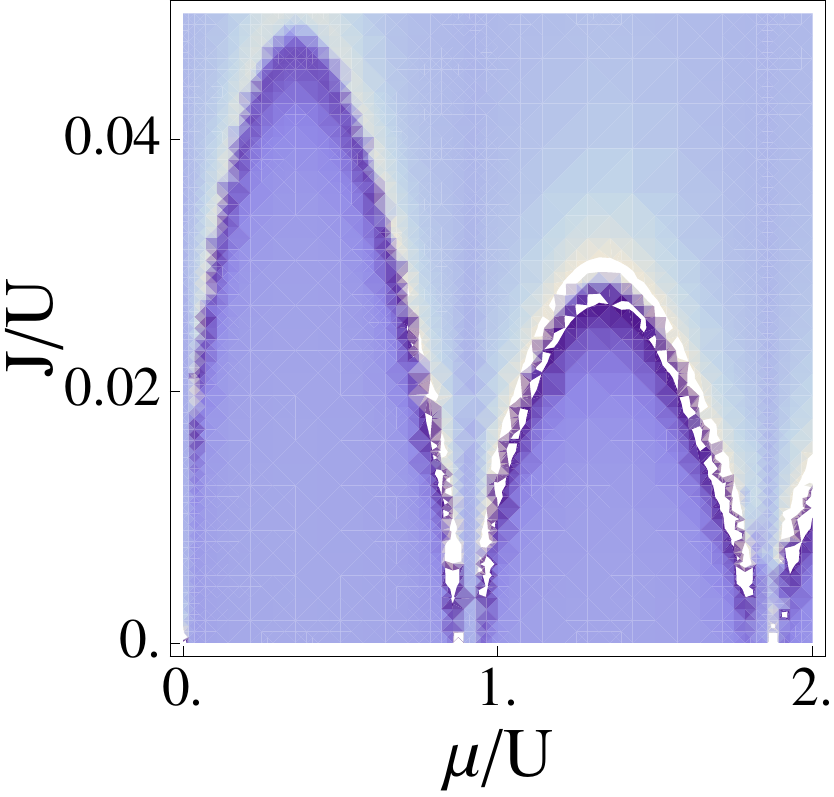}\label{fig:NQ005}}
\caption{\protect\footnotesize Numerical density plots of $\text{Re}\mathcal{G}^{K}_{00}$ for bath temperature $\beta U =25$, coupling width $\Lambda / U = 3$, and strength $\eta = 0.01$. \subref{fig:PDEQ} Equilibrium: $V=0$. \subref{fig:NQ05} Superfluid enhancement: $V/U =0.1$, $\hbar \Omega / U =0.05$, $\mathbf{k}=\frac{\pi}{a} \mathbf{x}$, $n_{p}=5$. \subref{fig:NQ005} Superfluid enhancement: $V/U =0.1$, $\hbar \Omega / U =0.005$, $\mathbf{k}=\frac{\pi}{a} \mathbf{x}$, $n_{p}=5$. Note the similarity to the $T=0$ equilibrium diagram. }
\label{PDfig}
\end{figure}

To determine the phase boundary, we are only interested in the stationary behavior of the system given by $\hat{\mathcal{G}}_{00}$. Inverting the block-diagonal $n_{p}^2 \times n_{p}^2$ matrix in Eq.\ (\ref{Dvec}), the real part of the Keldysh component of our correlation function, $\text{Re} \mathcal{G}^{K}_{00}$, can be displayed. Fig.\ \ref{fig:PDEQ} shows the system at equilibrium ($V=0$) including the effects of Ohmic dissipation. The phase boundary is given by the points where $\text{Re} \mathcal{G}^{K}_{00}$ diverges, and our results match those of Ref. \cite{Dalidovich09}. Fig.\ \ref{fig:NQ05} is an example of dynamic enhancement of superfluidity. Note the similarity of Fig.\ \ref{fig:NQ05} to the $T=0$ phase diagram at equilibrium. In Fig.\ \ref{fig:NQ005}, we see the effect of making the pulse energy $\hbar \Omega$ ten times smaller. Again there is superfluid enhancement, but notice that the valleys come to much finer points closer to the $\mu / U$ axis. This advocates an interpretation wherein our perturbation excites phase-stabilizing collective modes and artificially closes the energy gap. A smaller perturbation energy $\hbar \Omega$ is resonant with a smaller gap. Thus, the phase boundary is much closer to the $\mu / U$ axis at integer values of $\mu / U$ where the gap goes to zero. It becomes qualitatively identical to the $T=0$ equilibrium phase diagram.

\begin{figure}[t]
\vspace{0.186in}
\includegraphics[trim=1.4in 3.8in 0.5in 1.6in, clip=true, width=0.45\textwidth]{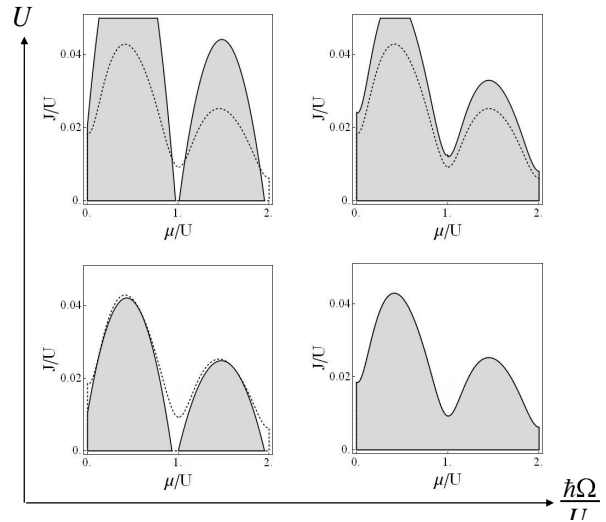}
\caption{{\protect\footnotesize The $T = 0.04 U$ equilibrium phase boundary (dotted line) plotted with examples from different regions of driving parameter space.}}
\label{PSmap}
\end{figure}

 The system's behavior in other regions of $V,\Omega$ space are catalogued in Fig.\ \ref{PSmap}. For instance, at high frequencies ($\hbar \Omega \gg U$), we see the familiar phenomenon of dynamic localization \cite{Zenesini10,Creffield06}. All Wigner components other than $N=0$ can be ignored, and the tunneling is renormalized $J \rightarrow J \mathcal{J}_{0}\left(V/\Omega \right)$ by the zeroth Bessel function of the first kind coming from the expansion of the non-equilibrium phase factor in Eq.\ (\ref{Hsub}). If $V/\Omega$ is tuned to a zero of $\mathcal{J}_{0}\left(x \right)$, we have dynamic suppression of tunneling. We find a featureless phase diagram (not shown in Fig.\ \ref{PSmap}) because there is no region of $J$ vs. $\mu$ space that admits long-range phase-coherence.

This phenomenon, familiar from driven Josephson arrays, can be understood as a cancellation of the dynamic phase acquired due to one period of driving with that of the hopping between sites. The bosons become localized with no long-range correlations, and $\text{Re} \mathcal{G}^{K}_{00}$ diverges nowhere. If we now tune $V$ toward zero, the phase boundary returns to its equilibrium form. If instead $\Omega$ is tuned lower, it will eventually be small enough to be resonant with the low energy modes available when $\mu / U$ is close to an integer. We will again have dynamic enhancement of superfluidity. However, further from the valley, there will be no available low-energy modes, and we will see suppression of tunneling. The result is larger Mott lobes that almost touch the $\mu / U$ axis. 

We have demonstrated the enhancement of the superfluid region in parameter space by driving. The experimental signature of this effect is similar to what has been found in time-of-flight experiments \cite{Greiner02}. When $\mu$ and $J$ are tuned to a point within the enhanced superfluid region close to integer values of $\mu/U$, there will be well-defined peaks in momentum space when the perturbation is on ($V \neq 0$) and a featureless interference pattern corresponding to destroyed phase coherence when the perturbation is off ($V=0$). 

The authors are grateful to Ehud Altman, William Phillips, and Hong Ling for illuminating discussions of this topic. This research was supported by JQI-PFC (A.R.), DARPA and US-ARO (V.G.), and the Packard Foundation, Sloan Foundation, and NSF grants PHY-0456720 and PHY-0803371 (G.R.)

\bibliographystyle{apsrev}
\bibliography{NEQBHM}

\end{document}